# Security bounds for efficient decoy-state quantum key distribution

Marco Lucamarini, James F. Dynes, Bernd Fröhlich, Zhiliang Yuan, and Andrew J. Shields

*Abstract*—Information-theoretical security of quantum key distribution (QKD) has been convincingly proven in recent years and remarkable experiments have shown the potential of QKD for real world applications. Due to its unique capability of combining high key rate and security in a realistic finite-size scenario, the efficient version of the BB84 QKD protocol endowed with decoy states has been subject of intensive research. Its recent experimental implementation finally demonstrated a secure key rate beyond 1 Mbps over a 50 km optical fiber. However the achieved rate holds under the restrictive assumption that the eavesdropper performs collective attacks. Here, we review the protocol and generalize its security. We exploit a map by Ahrens to rigorously upper bound the Hypergeometric distribution resulting from a general eavesdropping. Despite the extended applicability of the new protocol, its key rate is only marginally smaller than its predecessor in all cases of practical interest.

## I. INTRODUCTION

QUANTUM key distribution (QKD) [1], [2] in two decades has progressed considerably and reached a maturity suitable for real-world use. Fundamental achievements have been obtained in QKD theory and experiments [3]-[12]. On the theoretical side, security proofs have been extended beyond the "asymptotic scenario", accounting for the fact that real data samples are always finite and subject to statistical fluctuations [13]-[19]. This led to an operational definition of the security of QKD, aimed at quantifying through an $\varepsilon$-value the deviation of a real system from an ideal one. On the experimental side, QKD systems capable of achieving $\varepsilon$-values as small as $10^{-10}$ have been developed [20]-[24].

In order to bring QKD technology closer to real-world deployment, it is necessary to further reconcile the requirements of the theory with those of a real-world implementation, such as high key rate generation and low manufacturing costs. Therefore QKD protocols are continuously refined to approach the desired levels of efficiency and security.

Here, we review and extend a version of the efficient BB84 protocol [25]-[27] endowed with decoy states [28]-[31], recently introduced and experimentally realized in [24], which provides a key rate beyond 1 Mbps over a 50 km optical fiber with an $\varepsilon$-value of $10^{-10}$. This key rate was obtained under the limiting assumption that Eve performs collective attacks [1], [2]. In this case, the measured QKD quantities were represented by independent and identically distributed (i.i.d.) random variables that were bounded using the Clopper-Pearson (CP) confidence interval [32]-[34] for the Binomial distribution. By combining such bounds with the proof method described in [4], [7] and refined in [17]-[19], the security of the protocol was finally obtained.

Recently, a class of QKD protocols have been proven secure using the uncertainty relation for smooth entropies [35], [36]:

$$H_{\min}^{\varepsilon}(Z|E) + H_{\max}^{\varepsilon}(X|X') \geq \gamma n . \quad (1)$$

Eq. (1) holds if the transmitter is endowed with a perfect single photon source. The parameter $\gamma \in [0,1]$ is a quality factor related to the bias between the bases used by the transmitter [36]. If the emitted states are in two mutually unbiased bases, e.g. $Z$ and $X$, like in the ideal BB84 protocol, then $\gamma = 1$. The conditional smooth min entropy $H_{\min}^{\varepsilon}(Z|E)$ quantifies how many random bits are contained in $Z$ that are independent of Eve and $\varepsilon$-close to a uniform distribution, with $\varepsilon \geq 0$ the smoothing parameter [37]. $H_{\max}^{\varepsilon}(X|X')$, the conditional smooth max entropy, gives the number of additional bits necessary to reconstruct $X$ from $X'$ with failure probability $\varepsilon$. The key rate resulting from Eq. (1) is secure under general attacks so it can be used to drop the assumption of collective attacks from the efficient decoy-state BB84 protocol, as in [38].

However, additional work is required to guarantee security against the most general attack related to how the QKD quantities are sampled in a situation where the size of the sample is finite. The sampled quantities are random variables obeying a given distribution, in most cases Binomial, due to the two-valued nature of QKD observables. The Binomial distribution well represents experimental results under the i.i.d. assumption, or when measurements can be described as an operation of sampling *with* replacement. In some cases, however, this kind of sampling is not possible even in principle, for example, when sampling in the basis $X$ prevents sampling in the complementary basis $Z$, or vice versa [35], [39]. Under these circumstances, sampling *without* replacement has to be considered instead, and the Binomial distribution has to be replaced by the Hypergeometric distribution [35].

Below, we review the protocol of Ref. [24] and show its security under Eq. (1), along the lines described in [35] and [38]. We generalize the estimation procedure so as to cover both

Manuscript submitted August 1, 2014. This work has been partly supported by the Commissioned Research of National Institute of Information and Communications Technology (NICT), Japan.

Authors are with Toshiba Research Europe Ltd, 208 Cambridge Science Park, Cambridge, CB4 0GZ, United Kingdom. (E-mail: marco.lucamarini@crl.toshiba.co.uk; james.dynes@crl.toshiba.co.uk; bernd.frohlich@crl.toshiba.co.uk; zhiliang.yuan@crl.toshiba.co.uk; andrew.shields@crl.toshiba.co.uk).

M. Lucamarini, J. F. Dynes, Z. Yuan and A. J. Shields also with Toshiba Corporate Research & Development Center, 1 Komukai-Toshiba-Cho, Saiwai-ku, Kawasaki 212-8582, Japan.

the Binomial and the Hypergeometric distributions. This is done using a map by Ahrens described in [40]. It allows to reduce the general case to one that deals with Binomial distributions only. In turn, this allows to continue using the CP confidence interval for the Binomial distribution to provide worst-case bounds to the parameters of the protocol, as it was done in [24].

In Section II, we give some preliminary description of the Ahrens map and the CP confidence interval for the Binomial distribution. In Section III, we provide a detailed description of our protocol. In Section IV, we discuss the protocol security. Section V is left for the concluding remarks.

## II. Preliminaries

In the following, we give the basic notions about the Ahrens map and the CP confidence interval for the Binomial distribution. We will use them later for the protocol description and the security analysis.

### A. The Ahrens map

Consider a total population of $N$ balls in an urn containing $K$ white balls and $N - K$ black balls. A sample of $n$ elements ($n < N$) is drawn at random from the urn. A success is when a white ball is selected. If the sampled elements are not replaced in the urn, then the probability to draw $k$ white balls is given by the Hypergeometric distribution (HG):

$$\text{HG}(N, n, K, k) = \binom{K}{k}\binom{N-K}{n-k}/\binom{N}{n}, \quad (2)$$

which is positive for $\max(0, n - N + K) \leq k \leq \min(n, K)$. If the sampled elements are replaced in the urn, the probability of a successful event is constant, equal to $p = K/N$, and the probability to draw $k$ white balls from the urn is given by the Binomial distribution (BI):

$$\text{BI}(n, p, k) = \binom{n}{k} p^k (1-p)^{n-k}, \quad (3)$$

which is positive for $0 \leq k \leq n$.

The Ahrens map [40] is a permutation of the parameters $n, K, N - n, N - K$ so to obtain a new BI with the following property:

$$\text{HG}(N, n, K, k) \leq \sqrt{2}\, \text{BI}(\tilde{n}, \tilde{K}/N, \tilde{k}), \quad (4)$$

where the tilde indicates the permuted parameters, as defined by the following selection rules:

$$\tilde{n} = \min(n, K, N - n, N - K)$$
$$\text{IF } \tilde{n} = n \vee (N - n) \text{ THEN } \tilde{K} = \min(K, N - K)$$
$$\text{IF } \tilde{n} = K \vee (N - K) \text{ THEN } \tilde{K} = \min(n, N - n). \quad (5)$$

The permutation of the parameters is always possible, so there is no need to specify a range of application for it. In the top diagram of Fig. 1 we illustrate the Ahrens map, using a particular choice of the parameters. The curve $(i)$ is the distribution of $k$ according to $\text{BI}(n, K/N, k)$; the curve $(ii)$ is $\text{HG}(N, n, K, k)$; the curve $(iii)$ is the upper bound $\sqrt{2}\, \text{BI}(\tilde{n}, \tilde{K}/$ $N, \tilde{k})$ provided by the permuted BI distribution. The standard BI distribution has a larger variance than the corresponding HG, but it does not upper bound it on the whole range. On the contrary, the permuted BI distribution multiplied by $\sqrt{2}$ is always above the HG, so it can be used to upper bound it.

In some cases, the standard BI still provides bounds that are looser than those of the permuted BI. Our system automatically selects the loosest bounds, for each QKD session, so to guarantee the highest security level. This also simplifies the analysis because we only have to deal with BI distributions, either permuted or not.

### B. CP confidence interval

Consider a sequence of Bernoulli experiments in which the probability to obtain a success is constant, $p$. A sample of $n$ elements would then provide $k$ successes with the probability specified in Eq. (3).

Rather than obtaining the probability for $k$ successes, we are interested in confidence bounds for $k$, assessing that for any $\varepsilon > 0$, the true value of $k$ belongs to the interval $(\underline{k}, \overline{k})$ with confidence $\geq 1 - \varepsilon$, where $\underline{k}, \overline{k}$ are lower and upper bounds to the number of successes, respectively. This is obtained by solving in $p$ the following equations for the cumulative BI distribution [32], [41]:

$$P(k \leq \underline{k}) = \sum_{k=0}^{\underline{k}} \binom{n}{k} p^k (1-p)^{n-k} = \varepsilon, \quad (6)$$

$$P(k \geq \overline{k}) = \sum_{k=\overline{k}}^{n} \binom{n}{k} p^k (1-p)^{n-k} = \varepsilon. \quad (7)$$

The solutions of Eqs. (6) and (7) are respectively $\underline{p}$ and $\overline{p}$, and can be efficiently computed [41]. When the above equations are simultaneously solved, the resulting CP confidence interval contains $k$ with probability $1 - 2\varepsilon$. When the permuted BI is used to bound a HG distribution, Eqs. (6), (7) have to be solved with $\varepsilon/\sqrt{2}$ replacing $\varepsilon$, in order to obtain results with the same confidence. The system resets from $\varepsilon$ to $\varepsilon/\sqrt{2}$ automatically, if necessary.

In the bottom diagram of Fig. 1, we pictorially illustrate the lower bounds obtained though the CP approach, for the same probability distributions considered in the top diagram of Fig. 1. Lower bounds with confidence $1 - \varepsilon$ are given by the intersections of the cumulative functions with the line $\varepsilon$. In the example of the figure, the loosest bound is provided by the non-permuted BI distribution, labelled with $(i)$. So, in this case, our system would automatically select this bound to assess the security of the protocol. However, it is not always guaranteed that the non-permuted BI distribution upper bounds the HG distribution, labelled with $(ii)$. For that, we can use the upper bound provided by the Ahrens map, Eq. (4), labelled with $(iii)$.

## III. Protocol Description

In this Section, we modify the protocol described in Ref. [24] in order to generalize its security. In the following, we adopt a basis index $b = \{b_1, b_2\} = \{Z, X\}$ to indicate the bases chosen by the users, and a class index $\mu = \{\mu_1, \mu_2, \mu_3\} = \{u, v, w\}$ to indicate the intensity, or photon flux, used by the transmitter in preparing the light pulses. We denote "signal" ($u$), "decoy" ($v$) and "vacuum" ($w$) the three intensity classes used. Usually, $u > v > w \geq 0$. The basis will be chosen with probability $p_Z > p_X = 1 - p_Z$ and the class with probability $p_u > p_v \geq p_w = 1 - p_u - p_v$. We assume that the transmitter has a phase-randomised source of coherent states [42], [43] and that the intensity of the light pulses can be set with arbitrarily high precision. This makes the light source statistically equivalent to a Poissonian distribution of number states such that the probability to send a light pulse containing $k$ photons is $e^{-\mu}\mu^k/k!$. All the steps of the protocol and its final rate will be specified assuming the key bits are distilled only from the majority class $u$ and the majority basis $Z$. With minor modifications, key bits can be distilled from other classes and from $X$ basis too. This extra resource can be useful when the basis ratio $p_Z/p_X$ approaches 1 or when $p_u$, $p_v$ and $p_w$ have comparable magnitudes. The choice of a single basis is dictated by practicality considerations and is not necessary for security.

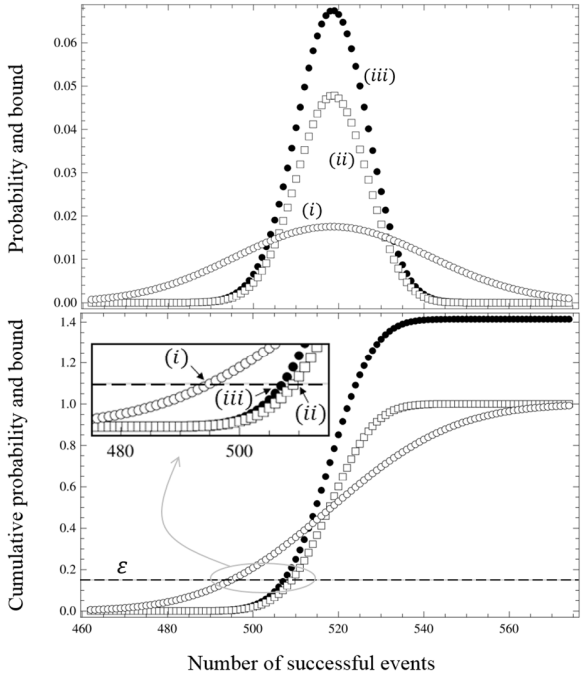

Fig. 1. Comparison between bounds used in the parameter estimation stage of QKD. Empty circles, label (i): binomial probability distribution BI($n, K/N, k$) (top) and corresponding cumulative distribution (bottom). Empty squares, label (ii): hypergeometric probability distribution HG($N, n, K, k$) (top) and corresponding cumulative distribution (bottom). Filled circles, label (iii): upper bound to the hypergeometric distribution (top) and to the corresponding cumulative distribution (bottom) by a recalibrated binomial distribution BI($\tilde{n}, \tilde{K}/N, \tilde{k}$) multiplied by $\sqrt{2}$. Inset: blow-up of the relevant points in the quantification of the security threshold, $\varepsilon$. The number of successes, $k$, is reported on the horizontal axis. Values used in the diagrams are: $N = 120{,}000$, $n = 103{,}820$, $K = 600$. Typical values in QKD are from 3 to 7 orders of magnitude greater.

*A. Transmitter*

With probability $p_\mu$, the transmitter (Alice) prepares a phase-randomised coherent state with intensity $\mu$. She then selects a basis $b$ with probability $p_b$ and a bit value 0 or 1 with probability 50%. She uses these values to encode a state that is sent to the receiver over the quantum channel.

*B. Receiver*

The receiver (Bob) chooses a basis $b$ with the same probability $p_b$ as Alice and then measures the incoming state using two threshold detectors $D_0$ and $D_1$. If no detector clicks, a vacuum count is recorded; if only detector $D_0$ ($D_1$) clicks, a bit value 0 (1) is recorded; if both detectors click, a random bit value, 0 or 1, is assigned and recorded [44], [45].

*C. Reconciliation and determination of samples size*

After a predetermined number of $N$ states have been sent by Alice and measured by Bob, users analyse the statistics associated to the states over an authenticated public channel. The very first time, the channel can be authenticated using a pre-shared secret string and universal$_2$ hashing [46]. Then, the secret string can be regenerated from the quantum key at every new session. At first, Bob discloses bases and timestamps of his non-vacuum counts. Then Alice announces bases and classes for these counts, together with the bit values in the $X$ basis and in the decoy and vacuum pulses. With these information, users form raw keys from all the counts in the class $\mu = u$ and matched bases $b_i = b_j = Z$, where $b_i$ and $b_j$ refer to Alice and Bob, respectively, and $i, j = \{1,2\}$. The length of the raw keys is denoted as $C_{uZZ}$. Similarly, the size of the set of non-vacuum counts with generic class and bases is denoted as $C_{\mu b_i b_j}$. The users can measure these quantities exactly.

From public communication, users can also compute the exact quantities $N_{\mu bb}$, i.e. the total number of pulses in the class $\mu$ and in the same basis $b_i = b_j = b$. In some cases, these quantities are very large and it is more practical to estimate upper and lower bounds for $N_{\mu bb}$ rather than determining the exact value on the classical channel. Due to the large size of the samples, the resulting bounds are tight and the confidence level very close to unity. To simplify the description, we omit the details of this issue in what follows and we just refer to the exact values $N_{\mu bb}$. The drawing of $C_{\mu bb}$ counts from $N_{\mu bb}$ pulses, in turn selected from a total population $N$, can give rise to a HG distribution, as first noted in [35]. As explained in Section II.A, the protocol automatically considers this possibility and, if necessary, treats it via Eq. (4). Because all the bits for $X$ basis, decoy and vacuum have been revealed, a direct comparison between Alice's and Bob's strings can tell the *exact* number of errors $E_{vZZ}$, $E_{wZZ}$ and $E_{\mu XX}$.

The users run a classical error correction (EC) algorithm to correct possible errors in the raw keys obtained from signals in the $Z$ basis. We call $E_{uZZ}$ the total number of errors in the raw keys and $n_{EC}$ the parity bits revealed in order to correct them. After EC, the users verify that error corrected keys are identical using universal$_2$ hashing. If the keys are found to be different, the protocol aborts and data are discarded. We call $\varepsilon_{\text{ver}}$ the probability that the keys are different but the protocol does not abort. In some cases, the verification step can be postponed until the authentication step, which is also performed using universal$_2$ hash functions. As a result of EC and verification, the users can estimate the number of errors $E_{uZZ}$ with confidence equal to or bigger than $1 - \varepsilon_{\text{ver}}$.

*D. Bounds to fluctuations and parameter estimation*

At this point, the quantities $N_{\mu bb}$, $C_{\mu bb}$ and $E_{\mu bb}$ are known to the users. They run the following steps to bound the finite-size fluctuations and estimate the unknown parameters of the protocol:

1. Bound yields and error rate in the minority basis using the CP confidence interval. Mean values and bounds are respectively $Y_b^{(\mu)} = C_{\mu bb}/N_{\mu bb}$, $\underline{Y}_b^{(\mu)}$, $\overline{Y}_b^{(\mu)}$ for the yields and $B_X^{(u)} = E_{uXX}/N_{uXX}$, $\overline{B}_X^{(u)}$ for the $X$ error rate. To obtain the bounds, the following distributions are considered. For the yields: $\text{BI}(N_{\mu bb}, Y_b^{(\mu)}, C_{\mu bb})$ and $\text{HG}(N, N_{\mu bb}, NY_b^{(\mu)}, C_{\mu bb})$; for the $X$ error rate: $\text{BI}(N_{uXX}, B_X^{(u)}, E_{uXX})$ and $\text{HG}(N, N_{uXX}, NB_X^{(u)}, E_{uXX})$. The HG distribution is bounded by the corresponding BI through the Ahrens map. Worst-case bounds are eventually selected, as described in Section II.

2. Numerical constrained optimization and decoy-state technique are combined with the bounds above to estimate $\underline{y}_b^{(k)}$ and $\overline{y}_b^{(k)}$, i.e. lower and upper bounds of the yields of the pulses containing $k$ photons ($k = \{0,1\}$) in the basis $b = \{Z, X\}$. Bounds to the number of $k$-photon pulses in the $b$ basis are then obtained as: $\underline{n}_b^{(k)} = \lfloor N_{ubb}\,\underline{y}_b^{(k)} e^{-u} u^k / k! \rfloor$, $\overline{n}_b^{(k)} = \lceil N_{ubb}\,\overline{y}_b^{(k)} e^{-u} u^k / k! \rceil$. The condition $\underline{n}_Z^{(1)} \gg \overline{n}_X^{(1)}$ is verified by the users, otherwise the protocol aborts. This condition is always met for $p_Z \gg p_X$.

3. In a similar way, the upper bound to the bit error rate of the 1-photon pulses in the $X$ basis, $\overline{q}_{\text{bit},X}^{(1)}$, is obtained. This is used as upper bound to the phase error rate in the $Z$ basis (see Section IV). If $\overline{q}_{\text{bit},X}^{(1)}$ is larger than a predetermined threshold $q_{\text{tol},X}$, protocol aborts. We call $p_{\text{abt}}$ the overall probability that the protocol aborts.

In Table I, we summarize all the quantities of the protocol together with the confidence level with which they are known, obtainable as the complement of the failure probability.

*E. Privacy amplification*

The users apply privacy amplification to their error corrected keys until they are left with the following number of bits:

$$n_{\text{sec}} \leq \underline{n}_Z^{(0)} + \gamma \underline{n}_Z^{(1)} - \overline{n}_Z^{(1)} h(q_{\text{tol},X}) - n_{\text{EC}} - \Delta . \quad (8)$$

All the quantities in the above rate equation have been previously defined, with the exception of $\Delta$, which amounts to:

$$\Delta = \log_2(2/\varepsilon_{\text{ver}}) + 6\log_2(46/\varepsilon_{\text{sec}}) , \quad (9)$$

TABLE I
QUANTITIES OF THE PROTOCOL

| Symbol | Quantity | Failure probability |
|---|---|---|
| $N$ | number predetermined triggers | exactly known, $\emptyset$ |
| $N_{\mu b}$ | sizes of samples in class $\mu$ and basis $b$ | exactly known to Alice, $\emptyset$ |
| $N_{\mu bb}$ | sizes of samples in class $\mu$ and matching bases $b$ | exactly known in principle estimated in practice, high confidence |
| $C_{\mu b_i b_j}$ | size of measured count samples | exactly known, $\emptyset$ |
| $\underline{Y}_b^{(\mu)}, \overline{Y}_b^{(\mu)}$ | bounds to the yields for the class $\mu$ | estimated, $2\varepsilon$ |
| $\underline{y}_b^{(k)}, \overline{y}_b^{(k)}$ | bounds to the yields of $k$-photon pulses in basis $b$ | estimated, $6\varepsilon$ |
| $\underline{n}_b^{(k)}, \overline{n}_b^{(k)}$ | bounds to number of $k$-photon pulses in basis $b$ | estimated, $6\varepsilon$ |
| $E_{uZZ}$ | errors in class $u$ and basis $Z$ | estimated, $\varepsilon_{\text{ver}}$ |
| $E_{\mu b_i b_j} \neq E_{uZZ}$ | errors in $\mu b_i b_j \neq uZZ$ | exactly known, $\emptyset$ |
| $\overline{B}_X^{(u)}$ | upper bound to $X, u$ BER | estimated, $\varepsilon$ |
| $\overline{q}_{\text{bit},X}^{(1)}$ | upper bound to $X$ QBER of 1-photon pulses | estimated, $19\varepsilon$ |
| $q_{\text{tol},X}$ | predetermined phase error | exactly known, $\emptyset$ |

Table I. Predetermined, measured and estimated quantities in the protocol, with their associated failure probability.

where $\varepsilon_{\text{sec}} = 10^{-10}$ defines the overall secrecy of the protocol[1]. The protocol is $\varepsilon_{\text{ver}} + \varepsilon_{\text{sec}}$ secure, meaning that it is $\varepsilon_{\text{ver}}$-correct and $\varepsilon_{\text{sec}}$-secret [35]. This definition of security is composable and allows to use the quantum key in cryptographic applications [37].

## IV. SECURITY

The security of the above protocol stems from two aspects. On one side, there is the estimation of Eve's information, quantified via the min-entropy [37], [38] and then upper bounded using the uncertainty principle [36], Eq. (1), and the max-entropy bound [47], [35]. On the other side, there is parameter estimation (PE). This is a refinement of the one adopted in [24]. However, we need to justify its application in this new context.

Let us start from a recap of what has been already achieved in terms of security for the efficient decoy-state BB84 protocol and compare it with our approach.

*A. State of the Art and Comparison*

In [19], the security of the efficient decoy-state BB84 (*eds-BB84*) protocol was initially demonstrated using the proof method in [17], [18], which holds under the assumption of collective attacks by Eve[2]. Due to non-optimized decoy-state bounds, the resulting performance in terms of key rate and working distance was quite poor. In [24], [49], the compatibility of the mentioned proof method with the CP approach and numerical PE was first demonstrated for the *eds*-BB84 protocol. This allowed to improve the decoy-state bounds and achieve experimental key rates beyond 1 Mbps over a 50 km optical fiber link, still under the condition of collective attacks [24]. In [50], the same numerical PE based on BI distribution as in [24] and [49] was used[3] to prove for the first time the security of the *eds*-BB84 protocol against general attacks, assuming a perfect vacuum state prepared by Alice and the quantities analogous to $N_{\mu bb}$ (see Table I) exactly known. The resulting HG distribution was upper bounded by the sum of two BI distributions [50]. Later on, a simpler proof of *eds*-BB84 security against general attacks, based on the entropic uncertainty relations [35], [36], was provided in [38]. In this case, the PE exploits Hoeffding's inequality [49], which is used to bound observable quantities analogous to $C_{\mu bb}$ and $E_{\mu bb}$ in Table I. Also, analytical expressions were used to estimate the parameters entering the key rate equation.

---

[1] The term $46\varepsilon$ in Eq. (9) is due to the use of $6 \times 3 + 19 = 37$ total constraints in the optimization problem, each of which can fail with probability $\varepsilon$, plus $9\varepsilon$ due to the proof method in [38].

[2] It was conjectured that the mentioned proof method holds for general attacks too, not only for collective ones. Recently, an attempt to prove this conjecture was made in [48] and it was found that a few extra bits have to be sacrificed during privacy amplification to go from collective to general attacks.

Here, we use the entropic uncertainty relations to quantify Eve's information and the CP confidence interval and numerical optimization to perform the PE. Differently from [50], we use the Ahrens map to tightly bound (within a factor $\sqrt{2}$, see Fig. 1) the HG distribution using a permuted BI distribution. This technique allows to always reduce the sampling from a HG distribution to one from a BI distribution. It is the first time the Ahrens map is used in QKD and we believe it represents a useful resource for the practical implementation. Moreover, we do not assume a perfect preparation of the vacuum state and the exact knowledge of the quantities $N_{\mu bb}$ (see Table I and Section III.C). Differently from [38], we use numerical optimization for PE. This provides tight bounds to the parameters, leading to a high key rate. As an indication, we obtain a key rate of 1.128 Mbps over 50 km of optical fiber (see Table II). With the same numerical parameters, a simulation of the protocol in [38] shows a key rate of 1.042 Mbps at 50 km, 7.5% lower. This is remarkable as our rate equation, Eq. (8), is more conservative than the one in [38], as the coefficient of $h(q_{\text{tol},X})$ in Eq. (8), $\overline{n}_Z^{(1)}$, is larger than the one in [38], $\underline{n}_Z^{(1)}$. Moreover, the key rate in our protocol is only due to the signal states sent in the $Z$ basis whereas all states and bases are used in [38].

### B. CP confidence interval and constrained optimization

As aid, in [35] Hoeffding's inequality [51], [52] and analytical expressions were used to upper bound the distance between the finite size value of certain quantities measured in QKD and their asymptotic values. For example, if $C$ counts are detected from a population of $N$ pulses prepared by Alice, the distance between the measured and the asymptotic values (labeled below with an asterisk) according to Hoeffding's inequality would be: $|C^* - C| \leq \sqrt{C/2 \ln(1/\varepsilon)}$, which holds with probability $1 - 2\varepsilon$.

Here, we do a similar operation using the CP method instead, applied to a (permuted or non-permuted) BI distribution, and numerical optimization, as explained in Section II.B. Specifically, given $C$ counts from $N$ pulses, the average detection probability is $Y = C/N$ and the bounds are $\underline{Y}, \overline{Y}$, obtained with confidence $1 - 2\varepsilon$ using the CP method. Hence, because $N$ is constant, we also have $|C^* - C| = N|Y^* - Y| \leq N(\overline{Y} - \underline{Y})$. The bounds $\underline{Y}_b^{(\mu)}, \overline{Y}_b^{(\mu)}$ and $\overline{B}_X^{(u)}$ in Table I are obtained in this way. The last one, $\overline{B}_X^{(u)}$, upper bounds the ratio $E_{uXX}/N_{uXX}$, i.e., the bit error rate (BER) in the $X$ basis.[4] These bounds are used, in turn, to estimate parameters that are not directly measurable, like $y_b^{(k)}$ and $q_{ph,Z}^{(1)}$. This is done through constrained optimization [53], as described in points $E.2$ and $E.3$ of the protocol. An example of optimization problem solved in our system is as follows [49]:

$$\min_\Gamma y_Z^{(1)}, \quad (10)$$

where $\Gamma$ is a set of constraints determined by: the measured quantities; the usual positivity and completeness conditions for probabilities; the following decoy-state QKD relations:

$$\underline{Y}_Z^{(\mu)} \leq e^{-\mu} \sum_k \frac{\mu^k}{k!} y_Z^{(k)} \leq \overline{Y}_Z^{(\mu)} \quad (\mu = \{u, v, w\}). \quad (11)$$

The optimization problem is linear and so efficiently solved. In the estimation of $y_Z^{(1)}$, three two-side nontrivial constraints are involved. Hence the overall $\varepsilon$-value for the simultaneous fulfillment of all constraints is conservatively bounded as $6\varepsilon$. With optimization problems similar to the one in Eqs. (10), (11), $k$-photon yields ($k = \{0,1\}$) in any basis can be obtained.

### C. Upper bound to the phase error rate

Numerical optimization is also used to upper bound the 1-photon quantum bit error rate (QBER) in the minority basis $X$ by solving the following problem:

$$\max_{\Gamma'} q_{\text{bit},X}^{(1)}, \quad (12)$$

where $\Gamma'$ contains the same constraints as for $y_Z^{(1)}$, plus the following one:

$$e^{-u} \sum_k \frac{u^k}{k!} y_X^{(k)} q_{\text{bit},X}^{(k)} \leq \overline{B}_X^{(u)}. \quad (13)$$

The above problem can be reduced to the following bound [49]:

$$q_{\text{bit},X}^{(1)} \leq \overline{q}_{\text{bit},X}^{(1)} = \left(e^u \overline{B}_X^{(u)} - \frac{1}{2} \underline{y}_X^{(0)}\right) \Big/ \left(u \, \underline{y}_X^{(1)}\right). \quad (14)$$

Nine two-side and one one-side nontrivial constraints are

---

For this reason we conservatively state that the proof method only guarantees security against collective attacks.

[3] See, e.g., Eq. (F.2) in [50], which is used to sample the Binomial distribution as in the Clopper-Pearson estimation method.

[4] Notice that this is different from the more common ratio $E_{uXX}/C_{uXX}$ known as "quantum bit error rate" (QBER) [33], [34].

involved in achieving this bound starting from the optimization problem in Eqs. (12) and (13). By weighting each of them with the same $\varepsilon$ value, we obtain that Eq. (14) holds with confidence $1 - 19\varepsilon$. To make the connection with security, we need to estimate the 1-photon phase error rate in the $Z$ basis, $q_{\text{ph},Z}^{(1)}$. For single photons, QKD theory guarantees that the asymptotic values of the QBER in the basis $X$ and the phase error rate in the basis $Z$ are the same:

$$q_{\text{bit},X}^{(1)*} = q_{\text{ph},Z}^{(1)*}. \qquad (15)$$

On the other side, the asymptotic value of a certain quantity coincides with its true value, and we know from the CP method that the QBER true value is bounded by $\overline{q}_{\text{bit},X}^{(1)}$ with confidence $1 - 19\varepsilon$. Therefore, the phase error rate is bounded by the same quantity with the same confidence:

$$q_{\text{ph},Z}^{(1)} \leq \overline{q}_{\text{bit},X}^{(1)}. \qquad (16)$$

Let us recall that this bound holds for both BI and HG distributions, because of the presence of the Ahrens map. In order to relate it to the security proof of [35] and [38], we need to add some details.

First, in Ref. [35] it is not the true value of the phase error rate to be used, but rather the bound to the phase error rate that a hypothetical observer would see if he tested a finite sample of size $n_Z^{(1)}$ in a population of $n_Z^{(1)} + n_X^{(1)}$ elements. Let us call $\tilde{q}_{\text{ph},Z}^{(1)}$ such a bound. We show here that $\overline{q}_{\text{bit},X}^{(1)}$ is a more conservative bound than $\tilde{q}_{\text{ph},Z}^{(1)}$, i.e., $\tilde{q}_{\text{ph},Z}^{(1)} \leq \overline{q}_{\text{bit},X}^{(1)}$. This implies that Eq. (16) still holds. Because $n_Z^{(1)} \geq \underline{n}_Z^{(1)} \gg \overline{n}_X^{(1)} \geq n_X^{(1)}$ (point III.D.2 of the protocol) we have that $\tilde{q}_{\text{ph},Z}^{(1)} \leq \tilde{q}_{\text{bit},X}^{(1)}$, where $\tilde{q}_{\text{bit},X}^{(1)}$ is estimated from $n_X^{(1)}$ single photons. We also have that $\tilde{q}_{\text{bit},X}^{(1)} \leq \overline{q}_{\text{bit},X}^{(1)}$ because, by negating this statement, we would obtain the absurd result that a bound estimated from a certain amount of coherent states via the decoy-state technique is tighter than one estimated directly from the same amount of single-photon states. This proves our statement.

Second, differently from [38], we keep the quality factor $\gamma$ of Eq. (1) in the estimation of the smooth min entropy via the uncertainty principle (compare with Appendix B in [38]). This leads to the factor $\gamma \underline{n}_Z^{(1)}$ in Eq. (8).

Third, we recalculate the bound to the smooth max-entropy according to the argument given in [35]. For that, we notice that all the steps in the supplementary materials of [35] can be repeated with the Serfling inequality [52] replaced by the CP confidence interval. In particular, the total number $W_{\text{ph},Z}$ of $Z$ phase errors can be bounded as:

$$W_{\text{ph},Z} \leq \lfloor \overline{n}_Z^{(1)} q_{\text{tol},X} \rfloor, \qquad (17)$$

with $q_{\text{tol},X}$ a predetermined threshold larger than $\overline{q}_{\text{bit},X}^{(1)} + O(1/\underline{n}_Z^{(1)})$. In turn, this implies that the smooth max-entropy is upper bounded by:

TABLE II
SECURE KEY RATE VERSUS DISTANCE

| Distance (km) | Key rate (bps) General attacks | Key rate (bps) Collective attacks |
|---|---|---|
| 30 | 3,124,188 | 3,413,432 |
| 50 | 1,128,172 | 1,251,857 |
| 70 | 364,787 | 414,334 |
| 90 | 82,997 | 98,112 |
| 110 | 1,448 | 1,589 |

Table II. Secure key rates versus optical fiber distance for the protocol of this work, secure against general attacks (column 2) and the one in [24], secure against collective attacks (column 3). In the new protocol, secure bits are distilled from the $Z$ basis only, while both $Z$ and $X$ bases contribute to them in [24]. For the simulation, the quality factor $\gamma$ has been set equal to 1 and optical fiber attenuation equal to 0.2 dB/km. $\varepsilon_{\text{sec}} = 10^{-10}$ and $\varepsilon_{\text{ver}} = 10^{-15}$. Detectors efficiency is 22.5%, afterpulse probability 5%, dark count probability/gate/detector $2.1 \times 10^{-5}$, number of detectors 2. Total insertion loss at receiver is 3dB. The acquisition time is 20 minutes. The values $p_X$, $p_\mu$, $\mu$ are optimized at every distance. At 50 km, they are: $p_X = 0.036$, $p_\mu = \{0.935, 0.028, 0.037\}$, $\mu = \{0.415, 0.05, 10^{-4}\}$, for the new protocol, and $p_X = 0.013$, $p_\mu = \{0.979, 0.011, 0.01\}$, $\mu = \{0.418, 0.03, 10^{-4}\}$, for the one in [24].

$$\overline{n}_Z^{(1)} h(q_{\text{tol},X}), \qquad (18)$$

where $h$ is the truncated binary entropy function. It could be worth remarking that Eq. (18) contains the upper bound to $n_Z^{(1)}$, $\overline{n}_Z^{(1)}$, which is clearly more conservative than the lower bound $\underline{n}_Z^{(1)}$ present in [38].

V. CONCLUSION

In this work, we extended the security proof of the efficient decoy-state BB84 protocol for QKD presented in [24] to cover the most general attack allowed by the laws of physics. We also added extra features to the protocol, like the possibility to drop the assumption of a perfect state preparation at Alice's side. This imperfection is included in the quality factor $\gamma$, which should be characterized by the users beforehand in a safe location.

Given the wider security range of the protocol, it is natural to ask whether its key rate is degraded respect to previous realizations. In Table II, we report values for the new protocol key rate versus optical fibre distance, and compare it with the protocol in [24], secure against collective attacks. At 50 km, the new protocol still provides beyond 1Mbps rate with 22.5% detection efficiency, well within the reach of current detectors [54]-[56]. Furthermore, the maximum achievable distance is more than 110 km. The new protocol compares well against the one in [24], whose key rate is recalculated and given in Table II, featuring on average only a 10% reduction.

The proof method in [35], adopted in our analysis, entails a reduced sensitivity to finite-size effects. The term $\Delta$ in Eq. (9) does not include the detrimental contribution proportional to the square root of the length of the raw key, $\sqrt{C_{uZZ}}$, which was present in [24]. In Fig. 2, we numerically simulate the secure key rate of the protocol (vertical axis) versus the size of the data block (top horizontal axis), which is varied by acting on the acquisition time (bottom horizontal axis). It can be seen that up to a block size of $10^5$, the key rate remains at more than 20% its asymptotic value. The minimum size of the sample providing a positive key rate is $1.6 \times 10^4$ bits.

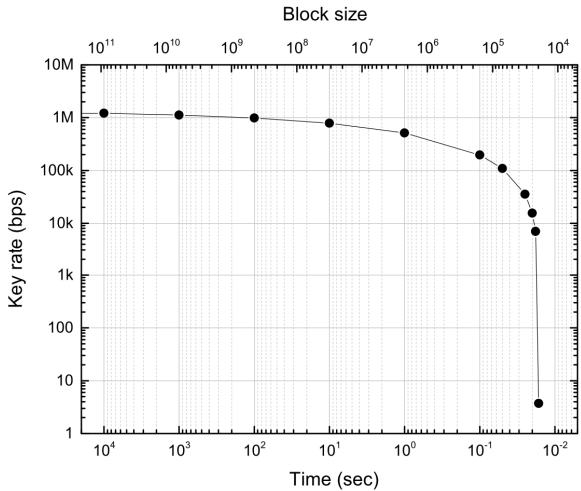

Fig. 2. Key rate versus acquisition time (bottom axis) and sifted block size (top axis), at a fixed optical fiber distance of 50 km. The smallest block size is $1.6 \times 10^4$, acquired in 16ms by a 1GHz-clocked system with the same parameters used to draw Table II. The values $p_X, p_\mu, \mu$ are optimized at every distance and at the smallest sample size they are: $p_X = 0.461$, $p_\mu = \{0.256, 0.392, 0.352\}$, $\mu = \{0.485, 0.097, 10^{-4}\}$. The quantity $q_{\text{tol}X}$ ranges from 3.4% in the asymptotic limit to 10% in the smallest size sample.

Overall, the performance of the here-presented decoy-state efficient BB84 protocol is comparable with what reported in the past [24], despite the wider class of attack covered in the new protocol and the single basis used to distill secure key bits. This is mainly due to the substantially unchanged numerical optimization in the parameter estimation stage. It still runs based on sampling from a Binomial distribution, which can be accomplished efficiently in several existing software packages. The gap between the Binomial and the Hypergeometric distributions, relevant for going from collective to general attacks, is bridged by the Ahrens map [40], that can be run automatically as a sub-routine of the numerical optimization program. We expect this to become a useful tool in other quantum communications protocols.